\RequirePackage{lineno}
\documentclass[aps,prl,twocolumn,showpacs,superscriptaddress,groupedaddress,nofootinbib]{revtex4}  
\usepackage{graphicx}  
\usepackage{dcolumn}   
\usepackage{bm}        
\usepackage{amssymb}   
\usepackage{ifpdf,latexsym,overpic,rotating}
\usepackage{array,multirow}
\usepackage{epsfig}
\usepackage[english]{babel}
\usepackage{amsmath}
\usepackage{color}   

\begin{document}

%

\newcommand{\HISKP}{Helmholtz-Institut f\"ur Strahlen- und Kernphysik, Universit\"at Bonn, Germany}
\newcommand{\PI}{Physikalisches Institut, Universit\"at Bonn, Germany}
\newcommand{\Juelich}{Institute~for~Advanced~Simulation,~Institut~f\"{u}r~Kernphysik
and J\"{u}lich~Center~for~Hadron~Physics, Forschungszentrum~J\"{u}lich,~D-52425~J\"{u}lich,~Germany}
\newcommand{\GIESSEN}{II. Physikalisches Institut, Universit\"at Giessen, Germany}
\newcommand{\BOCHUM}{Institut f\"ur Experimentalphysik I, Ruhr Universit\"at Bochum, Germany}
\newcommand{\BASEL}{Physikalisches Institut, Universit\"at Basel, Switzerland}
\newcommand{\GATCHINA}{National Research Centre "Kurchatov Institute", 
Petersburg Nuclear Physics Institute, Gatchina, Russia}
\newcommand{\FSU}{Department of Physics, Florida State University, Tallahassee, FL 32306, USA}

\title{Observation of the $\boldsymbol{p\eta'}$ Cusp in the New Precise Beam Asymmetry $\boldsymbol{\Sigma}$ Data for $\boldsymbol{\gamma p\to p \eta}$}
%
%
\affiliation{\HISKP}
\affiliation{\PI}
\affiliation{\Juelich}
\affiliation{\GIESSEN}
\affiliation{\BOCHUM}
\affiliation{\BASEL}
\affiliation{\GATCHINA}
\affiliation{\FSU}

\author{F.~Afzal}\email[Corresponding author: ]{afzal@hiskp.uni-bonn.de}\affiliation{\HISKP}
\author{Y.~Wunderlich}\affiliation{\HISKP}
\author{A.V.~Anisovich}\affiliation{\HISKP}\affiliation{\GATCHINA}
\author{D.~Bayadilov}\affiliation{\HISKP}\affiliation{\GATCHINA}
\author{R.~Beck}\affiliation{\HISKP}
\author{M.~Becker}\affiliation{\HISKP}
\author{E.~Blanke}\affiliation{\HISKP}
\author{K.-Th.~Brinkmann}\affiliation{\GIESSEN}
\author{S.~Ciupka}\affiliation{\HISKP}
\author{V.~Crede}\affiliation{\FSU}
\author{M.~Dieterle}\affiliation{\BASEL}
\author{H.~Dutz}\affiliation{\PI}
\author{D.~Elsner}\affiliation{\PI}
\author{S.~Friedrich}\affiliation{\GIESSEN}
\author{F.~Frommberger}\affiliation{\PI}
\author{A.~Gridnev}\affiliation{\GATCHINA}
\author{M.~Gottschall}\affiliation{\HISKP}
\author{M.~Gr\"uner}\affiliation{\HISKP}
\author{E.~Gutz}\affiliation{\GIESSEN}
\author{C.~Hammann}\affiliation{\HISKP}
\author{J.~Hannappel}\affiliation{\PI}
\author{J.~Hartmann}\affiliation{\HISKP}
\author{W.~Hillert}\affiliation{\PI}
\author{J.~Hoff}\affiliation{\HISKP}
\author{P.~Hoffmeister}\affiliation{\HISKP}
\author{C.~Honisch}\affiliation{\HISKP}
\author{T.~Jude}\affiliation{\PI}
\author{H.~Kalinowsky}\affiliation{\HISKP}
\author{F.~Kalischewski}\affiliation{\HISKP}
\author{I.~Keshelashvili}\affiliation{\BASEL}
\author{P.~Klassen}\affiliation{\HISKP}
\author{F.~Klein}\affiliation{\PI}
\author{E.~Klempt}\affiliation{\HISKP}
\author{K.~Koop}\affiliation{\HISKP}
\author{P.~Kroenert}\affiliation{\HISKP}
\author{B.~Krusche}\affiliation{\BASEL}
\author{M.~Lang}\affiliation{\HISKP}
\author{I.~Lopatin}\affiliation{\GATCHINA}
\author{P.~Mahlberg}\affiliation{\HISKP}
\author{U.-G.~Mei{\ss}ner}\affiliation{\HISKP}\affiliation{\Juelich}
\author{F.~Messi}\affiliation{\PI}
\author{V.~Metag}\affiliation{\GIESSEN}
\author{W.~Meyer}\affiliation{\BOCHUM}
\author{B.~Mitlas\'{o}czki}\affiliation{\HISKP}
\author{J.~M\"uller}\affiliation{\HISKP}
\author{J.~M\"ullers}\affiliation{\HISKP}
\author{M.~Nanova}\affiliation{\GIESSEN}
\author{K.~Nikonov}\affiliation{\HISKP}\affiliation{\GATCHINA}
\author{V.~Nikonov}\affiliation{\HISKP}\affiliation{\GATCHINA}
\author{V.~Novinskiy}\affiliation{\GATCHINA}
\author{R.~Novotny}\affiliation{\GIESSEN}
\author{D.~Piontek}\affiliation{\HISKP}
\author{G.~Reicherz}\affiliation{\BOCHUM}
\author{L.~Richter}\affiliation{\HISKP}
\author{D.~R\"onchen}\affiliation{\HISKP}
\author{T.~Rostomyan}\affiliation{\BASEL}
\author{B.~Salisbury}\affiliation{\HISKP}
\author{A.~Sarantsev}\affiliation{\HISKP}\affiliation{\GATCHINA}
\author{D.~Schaab}\affiliation{\HISKP}
\author{C.~Schmidt}\affiliation{\HISKP}
\author{H.~Schmieden}\affiliation{\PI}
\author{J.~Schultes}\affiliation{\HISKP}
\author{T.~Seifen}\affiliation{\HISKP}
\author{V.~Sokhoyan}\affiliation{\HISKP}
\author{C.~Sowa}\affiliation{\BOCHUM}
\author{K.~Spieker}\affiliation{\HISKP}
\author{N.~Stausberg}\affiliation{\HISKP}
\author{A.~Thiel}\affiliation{\HISKP}
\author{U.~Thoma}\affiliation{\HISKP}
\author{T.~Triffterer}\affiliation{\BOCHUM}
\author{M.~Urban}\affiliation{\HISKP}
\author{G.~Urff}\affiliation{\HISKP}
\author{H.~van~Pee}\affiliation{\HISKP}
\author{D.~Walther}\affiliation{\HISKP}
\author{Ch.~Wendel}\affiliation{\HISKP}
\author{U.~Wiedner}\affiliation{\BOCHUM}
\author{A.~Wilson}\affiliation{\HISKP}\affiliation{\FSU}
\author{A.~Winnebeck}\affiliation{\HISKP}
\author{L.~Witthauer}\affiliation{\BASEL}

\collaboration{The CBELSA/TAPS Collaboration}
 \vskip 0.25cm

%

\date{\today}

\begin{abstract}
Data on the beam asymmetry $\Sigma$ in the photoproduction of $\eta$ mesons off protons are reported for tagged photon energies from 1130 to 1790 MeV (mass range from $W=$ 1748 MeV to $W=$ 2045 MeV). The data cover the full solid angle that allows for a precise moment analysis. For the first time, a strong cusp effect in a polarization observable has been observed that is an effect of a branch-point singularity at the $p\eta'$ threshold [$E_\gamma=$ 1447 MeV ($W=$ 1896 MeV)]. The latest BnGa partial wave analysis includes the new beam asymmetry data and yields a strong indication for the $N(1895)\frac{1}{2}^-$ nucleon resonance, demonstrating the importance of including all singularities for a correct determination of partial waves and resonance parameters.
\end{abstract}

\maketitle
\noindent
Mesons and baryons are color neutral bound systems containing quarks and gluons. Understanding their dynamics within these bound systems and thereby also the generation of the hadron excitation spectrum still remains a challenging task since, in this regime, quantum chromodynamics (QCD), which is the field theory of strong interactions, cannot be solved perturbatively. In the baryon spectroscopy sector \cite{Klempt,Ireland2020,MeissnerNStar}, intense experimental efforts have been conducted in the last decade at different facilities like the Grenoble Anneau Acc\'{e}l\'{e}rateur Laser (GRAAL), the Electron Stretcher Accelerator (ELSA), the Mainz Microtron (MAMI) and Jefferson Lab that help to challenge quark models \cite{quarkmodel1,quarkmodel2,Loering}, lattice QCD calculations \cite{lattice}, and approaches based on effective field theories \cite{eft}.
To extract the contributing resonances from the measured data, partial wave analyses (PWAs) need to be performed. Different approaches such as a K matrix or N/D approach (e.g. BnGa PWA \cite{BnGa201402,etapaper,BnGa2017}), an isobar model [e.g. $\eta$MAID (Mainz Unitary Isobar Model) \cite{EtaMAID,Tiator:2018heh}], or a dynamical coupled-channel model [e.g. the J\"ulich-Bonn (J\"uBo) model \cite{JuBo15,JuBo17}] exist to address this problem. To constrain the present ambiguities of the partial waves, a large database is needed \cite{Wunderlich:2014xya,Beck:2016hcy} that comprises not only unpolarized cross sections but also single and several double polarization observables. In particular, the measurement of polarization observables in photon induced reactions is important due to their sensitivity to the interference of partial waves \cite{LFitPaper}. At the moment, a very good database exists only for the reaction $\gamma p \to p\pi^0$, while data are scarce for the $p\eta$ and the $p\eta'$ final states. However, high precision and a full angular coverage are important, especially in order to gain information for orbital angular momentum of the meson $\ell>3$ (higher than $F$ waves), which cannot be ignored for masses higher than $W\approx$ 2000 MeV. 

The partial wave amplitudes have to be known as functions of the energy in order to extract resonance parameters within a PWA. More specifically, these amplitudes have to be \textit{analytic} functions in the complex energy plane \cite{AnalyticSMatrix}. It is a well-known mathematical fact that analytic functions are uniquely specified by their singularities \cite{AnalyticSMatrix,BehnkeSommer}. The most relevant type of singularities is resonance poles \cite{AnalyticSMatrix, Ceci:2011ae}. The other important type of singularities is branch points. Every new channel opening introduces a branch point on the real energy axis at the threshold energy $s=\left( \sum_{i} m_{i} \right)^{2}$ \cite{AnalyticSMatrix}, where $m_i$ is the mass of each final state particle. This type of branch point, due to stable asymptotic states, can lead to a pronounced "cusp" effect, which is visible most prominently in the $S$ wave. 

The amplitude can only be correctly specified once all singularities are known and incorporated correctly. The absence of certain singularities may (greatly) influence the extracted location and properties of others. This is due to the fact that a fit with a physically incorrect amplitude, for instance one where an important cusp effect is not taken into account, will still try to describe the data on the real energy axis. Thus, the properties of the resonance poles could be sensitively disturbed (see e.g., the work by Althoff \textit{et al.} \cite{Althoff} where the $p\eta$ cusp was observed in the $p\pi^0$ and $n\pi^+$ cross section data for the first time).

In this Letter we report new data on the beam asymmetry $\Sigma$ for $\eta$ photoproduction in the incident-photon energy range of $E_\gamma=1130$ MeV to $E_\gamma=1790$\,MeV ($W=$ 1748 MeV to $W=$ 2045 MeV) and for an angular range of $-0.92<\cos\theta_\eta^{\text{cm}}<+0.92$. This wide angular range allows one to deduce the interference between the $S$ wave and $G$ wave in the extracted Legendre moments \cite{LFitPaper}. This term requires an $S$-wave resonance and a significant cusp at the $p\eta'$ threshold, which confirms the observation of the $p\eta'$ cusp reported by Kashevarov \textit{et al.} \cite{Kashevarov:2017kqb} in the total cross section for $\eta$ photoproduction. \newline

The beam asymmetry data were taken with the CBELSA/TAPS experiment using an electron beam of 3.2~GeV energy provided by ELSA \cite{ELSA}. The electron beam was scattered off a 500-$\mu$m-thick diamond crystal, producing linearly polarized photons via coherent bremsstrahlung \cite{LinPol}. 
The data were divided into two subsets using two different coherent edge positions at $E_\gamma\approx$ 1750~MeV and $E_\gamma\approx$ 1850~MeV with maximum polarization degrees of $\delta_l\approx$ 40\% at 1680~MeV and $\delta_l\approx$ 35\% at 1780~MeV, respectively. The linearly polarized photons were incident on a 5-cm-long liquid hydrogen target, which was located at the center of the Crystal Barrel calorimeter \cite{CB}. At the forward laboratory angles, the Crystal Barrel calorimeter was complemented by the Forward Plug and the MiniTAPS \cite{TAPS} calorimeters. All three electromagnetic calorimeters together covered almost the complete $4\pi$ solid angle and were highly efficient at detecting photons. Charged particles were identified by scintillating plates mounted in front of the Forward Plug and the MiniTAPS crystals, and by the inner scintillating fiber detector that surrounded the target. Detailed information on the experimental setup can be found in \cite{langesEpaper}. 

The decay mode $\eta \to \gamma \gamma$ with a branching ratio of 39.4\% \cite{PDG} was chosen to reconstruct the $\eta$ mesons in the reaction $\gamma p \to p\eta$. Therefore, at least two hits were required in the calorimeters. 
Optionally, a third hit was accepted as well that was detected either in one of the calorimeters or, in order to take low energetic protons into account, only in one of the charge sensitive detectors. 
If a third hit was detected in the calorimeters, all possible combinations were taken into account to assign two out of three hits to the two decay photons of the $\eta$. A $\pm2\sigma$ cut was applied to the invariant mass of the two decay photons around the $\eta$ mass. In addition, the proton was treated as a missing particle and a $\pm1.7\sigma$ cut was applied around the calculated missing proton mass. If a third hit was detected, the azimuthal angles of the meson and the proton had to form an angle of $180^\circ$ within $\pm 2 \sigma$. Furthermore, an angular correlation of the measured and calculated missing particle's polar angle within $\pm2\sigma$ was demanded. The majority of the background contributions stem from the $p\pi^0$ final state that arises due to a misidentification of the proton as one of the $\eta$ decay photons. They were almost entirely suppressed using the cluster properties of the final state particles. The combinatorial background is negligibly small. More details are given in \cite{FarahPhD}. In total, $5.24\times 10^5$ $p\eta$ events were selected with a low total background contribution of 2\%-3\% for a large angular range. At the extreme angles ($\cos\theta_{\eta}^{\text{cm}}<-0.7$ and $\cos\theta_{\eta}^{\text{cm}}>0.7$) the background contribution was found to be at most $6\%$.
The beam asymmetry values were corrected for this background contribution, considering the possibility of a nonzero background beam asymmetry in the systematic error as described in \cite{2pi0paper}.

Using only a linearly polarized photon beam the polarized cross section is given by \cite{Barkerpaper}
\begin{equation}
\label{equ:polcs}
\frac{d\sigma}{d\Omega}= \frac{d\sigma_0}{d\Omega}\left[ 1 - \delta_l \Sigma \cos\left(2\alpha - 2\varphi\right)\right]
\end{equation}
\noindent
where $\frac{d\sigma_0}{d\Omega}$ is the unpolarized cross section, $\alpha$ is the azimuthal angle of the polarization vector and $\varphi$ the azimuthal angle of $\eta$ in the laboratory frame. To extract the beam asymmetry $\Sigma$, event-based maximum likelihood fits were performed after binning the data only in the kinematic variables $E_\gamma$ and $\cos\theta_{\eta}^{\text{cm}}$. The azimuthal distribution of the events was described by 
\begin{equation}
\label{equ:pdfallgemein}
f_{\text{phy}}=\frac{d\sigma}{d\Omega}/\frac{d\sigma_0}{d\Omega} = 1 - \delta_l \Sigma \cos\left(2\alpha - 2\varphi\right).
\end{equation}
Possible detector asymmetries and random-time background in the selected data were addressed as described in detail in Ref. \cite{FarahPhD,2pi0paper} before minimizing the log-likelihood function.

\begin{figure}[htb]
\includegraphics[width=\columnwidth]{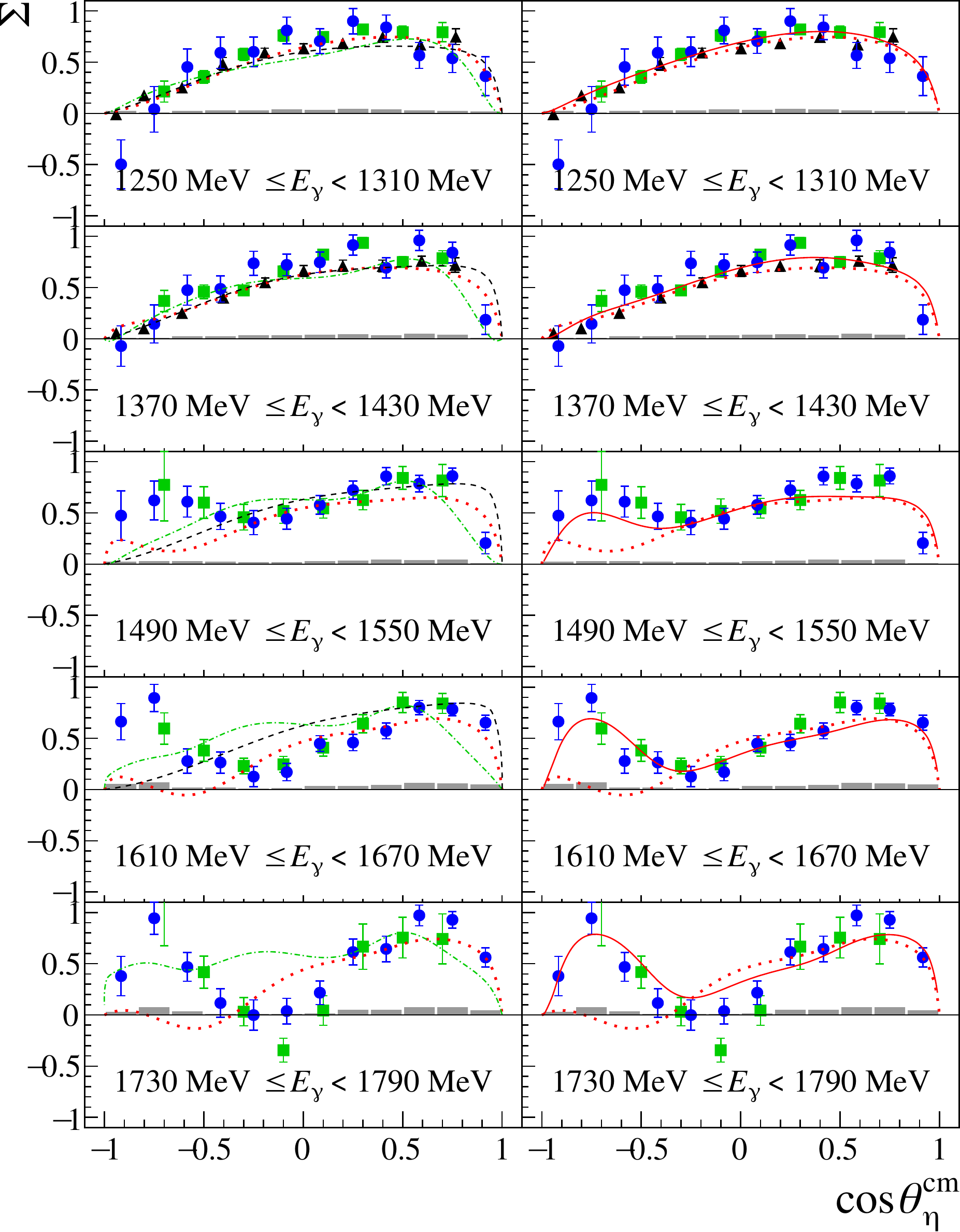}
\caption{New beam asymmetry $\Sigma$ data (blue points) shown as a function of $\cos\theta_{\eta}^{\text{cm}}$ for five different energy bins. The systematic uncertainties are given by the gray area. In the left-hand column, the new $\Sigma$ data are compared to existing data from the GRAAL collaboration (black triangles) \cite{GRAALdaten} and from the CLAS collaboration (green squares) \cite{CLASdaten}, and with different PWA predictions [BnGa-2014-02 (dotted red line) \cite{BnGa201402}, J\"uBo-2015-FitB (dash-dotted green line) \cite{JuBo15}, and $\eta$MAID-2003 (dashed black line) \cite{EtaMAID}]. In the right-hand column, the data are compared to the BnGa-2014-02 (dotted red line) \cite{BnGa201402} and their latest solution BnGa-2019 (solid red line) \cite{etapaper}, which includes the new $\Sigma$ data.}
\label{fig:beamasymbins}
\end{figure}

Figure~\ref{fig:beamasymbins} shows the beam asymmetry $\Sigma$ as a function of $\cos\theta_{\eta}^{\text{cm}}$ for five different energy ranges, two below and three above the $p\eta'$ photoproduction threshold at $E_\gamma=$ 1447 MeV ($W=$ 1896 MeV). The error bars contain only statistical uncertainties. The systematic uncertainty is depicted as well, which takes into consideration the aforementioned background contribution and the estimated uncertainty of the degree of linear polarization (5\%-8\%).
In the left-hand column, our new data are compared to existing data from GRAAL \cite{GRAALdaten} and the CEBAF Large Acceptance Spectrometer (CLAS) \cite{CLASdaten} and to different PWA predictions. The formation of a backward-angle peak above the $p\eta'$ threshold is remarkable, which the different PWAs do not predict at all. In the right-hand column, the new BnGa fit (BnGa-2019) is shown. It includes our new beam asymmetry data, the beam asymmetry CLAS data \cite{CLASdaten}, the differential cross section of the $p\eta$ and the $p\eta'$ channel of the A2 collaboration \cite{Kashevarov:2017kqb}, as well as the $E,G,T,P$, and $H$ CBELSA/TAPS data \cite{etapaper}.

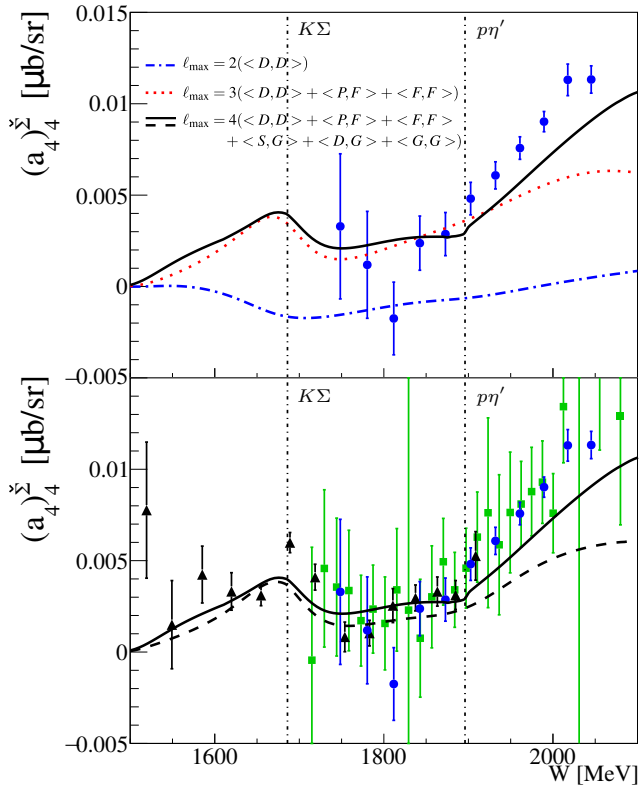
\begin{figure}
\begin{overpic}[width=\columnwidth]{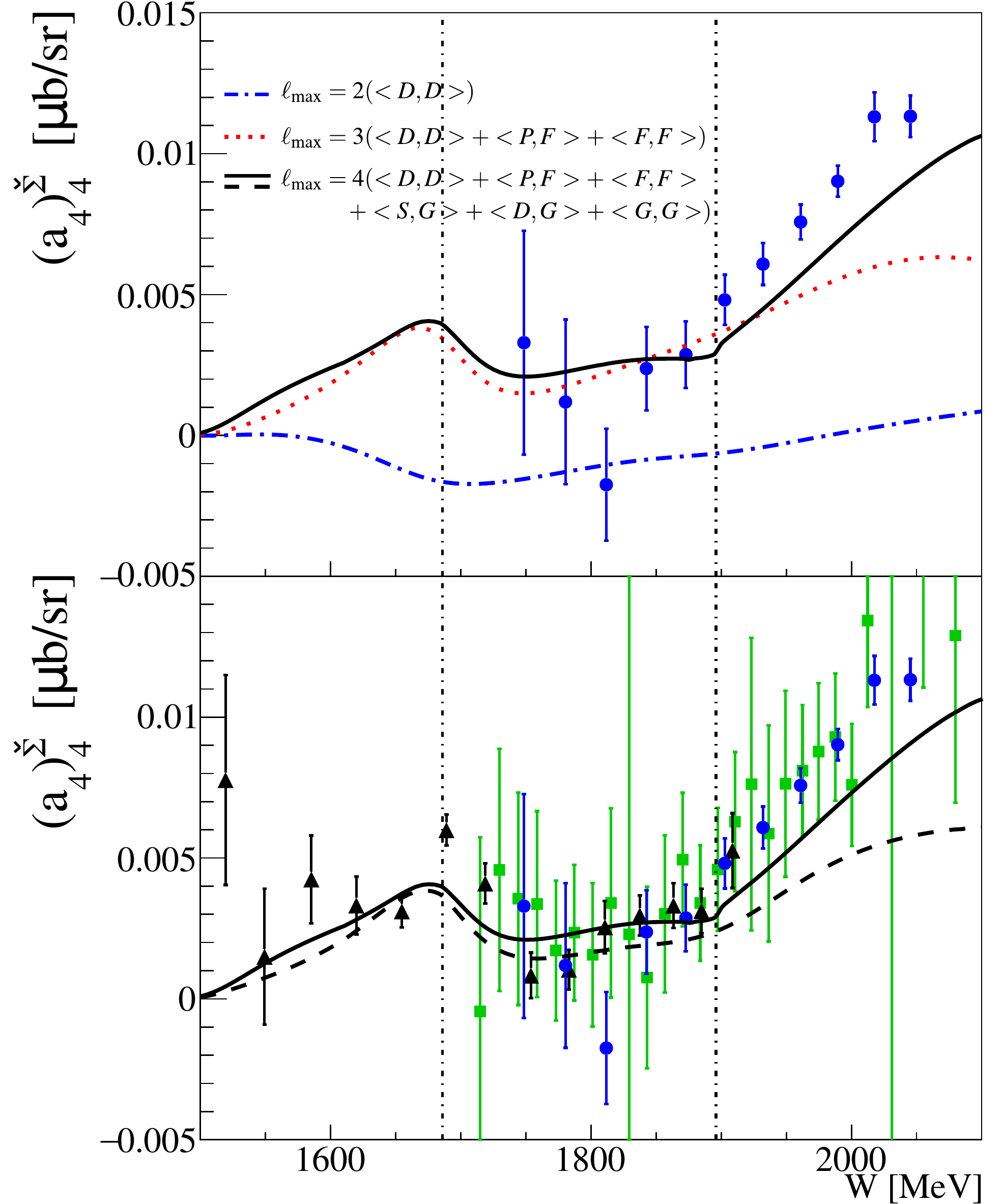}
\put(61,96){\scriptsize{$p\eta'$}}
\put(38,96){\scriptsize{$K\Sigma$}}
\put(61,49){\scriptsize{$p\eta'$}}
\put(38,49){\scriptsize{$K\Sigma$}}
\end{overpic}
\caption{Top: The energy dependence of the fit coefficient $(a_4)_4^{\check{\Sigma}}$ is shown for the $p\eta$ final state (blue points) as extracted from the new CBELSA/TAPS data according to Eq. (\ref{eq:SigmaAsymmetryPolynomialExp}). The continuous curves are evaluated using the BnGa-2019 PWA \cite{etapaper} and truncating at different $\ell_{\text{max}}$. Bottom: The GRAAL (black triangles) \cite{GRAALdaten} and the CLAS beam asymmetry data (green squares) \cite{CLASdaten} were analyzed according to Eq. (\ref{eq:SigmaAsymmetryPolynomialExp}), and the results for the fit coefficient $(a_4)_4^{\check{\Sigma}}$ are compared to the ones of the new CBELSA/TAPS data (blue points). The curve for $\ell_{\text{max}}$ = 4 is shown for the BnGa-2014-02 PWA \cite{BnGa201402} (dashed black line) and the BnGa-2019 PWA \cite{etapaper} (solid black line). }
\label{fig:beamasymLfits}
\end{figure}

To better understand the origin of the backward peak, we use the method of \textit{moment analysis} \cite{LFitPaper}. Truncating the well-known expansion of the photoproduction amplitude to electric and magnetic multipoles $\left\{ E_{\ell \pm}, M_{\ell \pm} \right\}$ at a maximal orbital angular momentum quantum number $\ell_{\mathrm{max}}$,
allows us to express the observables with finite polynomials in the angular variable $\cos \theta_\eta^{\text{cm}}$.
In Ref. \cite{LFitPaper}, these finite expansions are given in terms of associated Legendre polynomials $P_{\ell}^{m} \left( \cos \theta \right)$ \cite{Abramowitz}. The profile function $\check{\Sigma} = \sigma_{0} \hspace*{1pt} \Sigma$ belonging to the beam asymmetry $\Sigma$ reads
\begin{equation}
 \check{\Sigma} \left( W, \theta \right) = \frac{q}{k} \sum_{j = 2}^{2 \ell_{\mathrm{max}}} \left( a_{\ell_{\mathrm{max}}} \right)^{\check{\Sigma}}_{j} \left( W \right) \hspace*{1pt}  P^{2}_{j}  \left( \cos \theta \right)  \mathrm{.}  \label{eq:SigmaAsymmetryPolynomialExp}
\end{equation}
This expansion can, when fitted for different ascending quantum numbers, e.g., $\ell_{\mathrm{max}} = 1,2,3,$ and so on, be used as a test for the optimal order of partial waves $\ell_{\mathrm{max}}$ needed to describe the angular distribution \cite{LFitPaper}. \newline
The Legendre coefficients $\left( a_{\ell_{\mathrm{max}}} \right)^{\check{\Sigma}}_{j}$ themselves are bilinear hermitian forms of the multipoles $\left\{ E_{\ell \pm}, M_{\ell \pm} \right\}$. Using a short notation introduced in Ref. \cite{LFitPaper}, we write here the coefficient $\left(a_{4}\right)^{\check{\Sigma}}_{4}$ as an example:
\begin{align}
  \left(a_{4}\right)^{\check{\Sigma}}_{4} = & \left<D,D\right>  \nonumber \\
                    + &\left<P,F\right> + \left<F,F\right> \nonumber \\
                    + &\left<S,G\right> + \left<D,G\right>   + \left<G,G\right> \mathrm{.} \label{eq:SigmaCoeff4Composition}
\end{align}
The symbols $\left< -, - \right>$ are shorthand for a sum of bilinear products of multipoles, multiplying only waves of certain orbital angular momentum quantum numbers. We use here the spectroscopic notation, i.e., $\ell = 0,1,2,3,$ and so on corresponding to $S, P, D, F,$ and so on waves. \newline 
Using the method of moment analysis we found that $\ell_{\mathrm{max}} = 4$ is sufficient to obtain a satisfactory fit of the angular distributions of our data for $\check{\Sigma}$ [cf.\cite{BnGa201402} Eq.~\eqref{eq:SigmaAsymmetryPolynomialExp}], considering the $\chi^2/\text{ndf}$ values. The fit results remain stable against an increase from $\ell_{\mathrm{max}} = 4$ to $\ell_{\mathrm{max}} = 5$. Furthermore, an interesting interpretation of the backward peak arises once the extracted Legendre coefficients are studied. The results for the coefficient $\left(a_{4}\right)^{\check{\Sigma}}_{4}$, which is most sensitive to the $\left<S,G\right>$ interference, are shown in Fig.~\ref{fig:beamasymLfits}. 

\noindent
In the lowest five energy bins, this coefficient remains approximately constant at a value close to zero. Then, at an energy that corresponds precisely to the production threshold for the $p \eta^{\prime}$ final state (see Fig.~\ref{fig:beamasymLfits}), the fit results for $\left(a_{4}\right)^{\check{\Sigma}}_{4}$ show a noticeable and almost linear rise toward positive values. 
We applied the moment analysis also to the GRAAL and the CLAS beam asymmetry data (see Fig.~\ref{fig:beamasymLfits}). While the GRAAL data stop at around the $p\eta^{\prime}$ threshold, the CLAS data confirm the rise in the Legendre moment
$\left(a_{4}\right)^{\check{\Sigma}}_{4}$. However, the lack of precise backward and forward beam asymmetry CLAS data leads to much larger uncertainties.

The rise in the moment $\left(a_{4}\right)^{\check{\Sigma}}_{4}$ yields a consistent explanation for the emergence of the backward peak. First, the Legendre moment $\left(a_{4}\right)^{\check{\Sigma}}_{4}$ is multiplied by the polynomial $P^{2}_{4} \left( \cos \theta \right)$ in the expansion, Eq.~\eqref{eq:SigmaAsymmetryPolynomialExp}, and $P^{2}_{4}$, is the \textit{lowest} associated Legendre polynomial $P^{2}_{\ell}$ which has a pronounced forward \textit{and} backward peak. Thus, the polynomial $P^{2}_{4}$ is at least required to describe an angular distribution with a backward peak. Second, for the first time, the $p \eta^{\prime}$ cusp has been observed directly in a Legendre moment of a polarization observable. Since only the $S$ wave $E_{0+}$ contains a cusp at all, we can see that the most important contribution producing the cusp in $\left(a_{4}\right)^{\check{\Sigma}}_{4}$ arises due to the $\left< S, G \right>$ interference term in Eq.~\eqref{eq:SigmaCoeff4Composition}. Thus, for energies below the $p \eta^{\prime}$ threshold, $\left(a_{4}\right)^{\check{\Sigma}}_{4}$ is consistent with zero (Fig.~\ref{fig:beamasymLfits}) and no backward peak arises (Fig.~\ref{fig:beamasymbins}), whereas above the $p \eta^{\prime}$ threshold the backward peak can be seen in the $\Sigma$ angular distributions (Fig.~\ref{fig:beamasymbins}). Correspondingly, the cusp effect has enhanced the moment $\left(a_{4}\right)^{\check{\Sigma}}_{4}$ (Fig.~\ref{fig:beamasymLfits}). 

\begin{figure}
\begin{overpic}[width=\columnwidth]{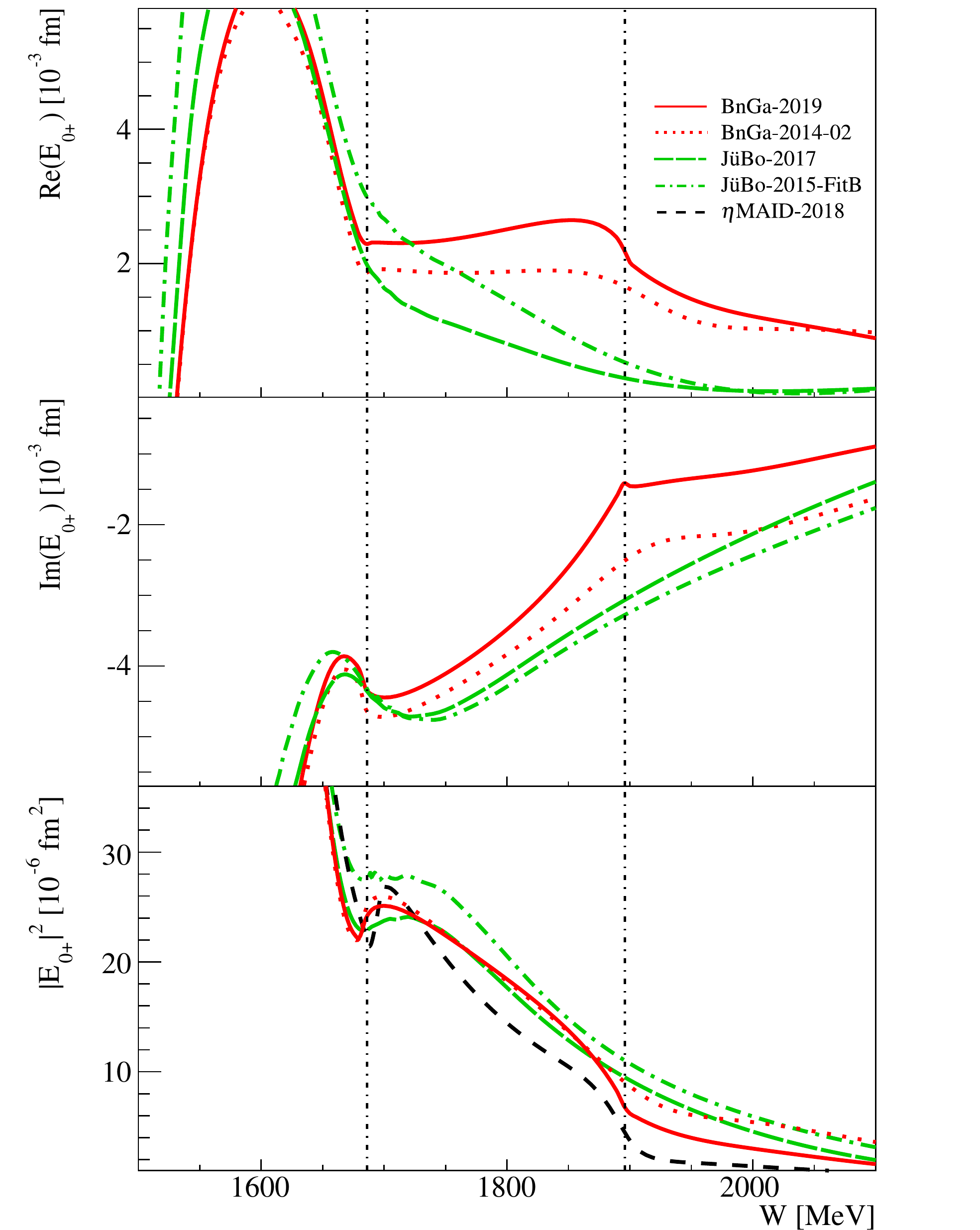}
\put(52,33){\scriptsize{$p\eta'$}}
\put(31,33){\scriptsize{$K\Sigma$}}
\put(52,64.5){\scriptsize{$p\eta'$}}
\put(31,64.5){\scriptsize{$K\Sigma$}}
\put(52,96){\scriptsize{$p\eta'$}}
\put(31,96){\scriptsize{$K\Sigma$}}
\end{overpic}
\caption{The real part (upper plot), and imaginary part (middle plot), and absolute square (lower plot) of the $E_{0+}$ multipole are depicted for the $p\eta$ final state for several different PWA. The imaginary and real parts are not shown for the $\eta$MAID-2018 PWA due to different phase conventions \cite{Tiator:2018heh}. 
}
\label{fig:multipole}
\end{figure}
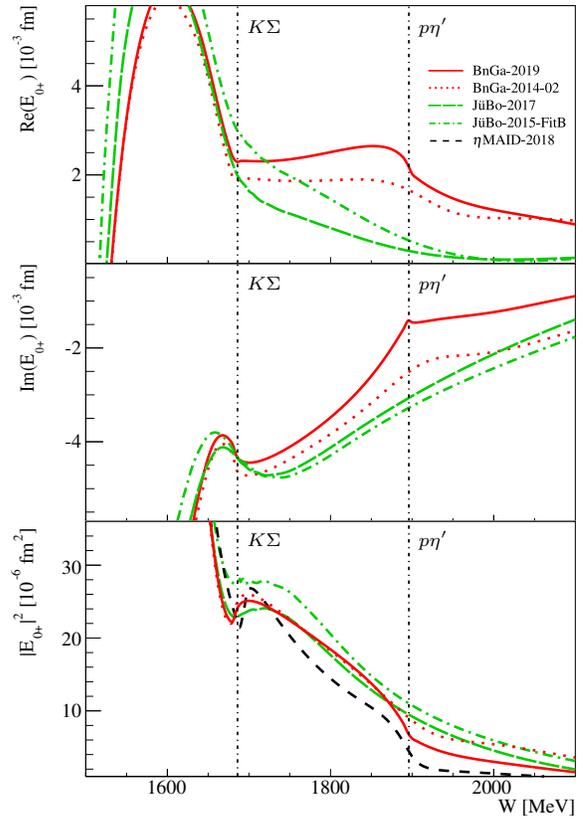

Our claim, that an effect of the cusp structure in the $p\eta$ $S$ wave ($E_{0+}$ multipole) is observed at the $\eta^{\prime}$ threshold, is substantiated by a comparison of the fitted Legendre moment $\left(a_{4}\right)^{\check{\Sigma}}_{4}$ to the BnGa PWA. In Fig.~\ref{fig:beamasymLfits}, the evaluation of the coefficient $\left(a_{4}\right)^{\check{\Sigma}}_{4}$ using $\eta$ photoproduction multipoles from the BnGa PWA is shown for two solutions, namely the BnGa-2014-02 (prediction) \cite{BnGa201402} and a more recent BnGa-2019 (fit solution including the new $\Sigma$ data) \cite{etapaper}. While in the more recent solution, the $\eta^{\prime}$ cusp has been included properly to achieve a good description of the new $\Sigma$ data and the total cross section data from A2-MAMI \cite{Kashevarov:2017kqb} which is most prominently visible in the $S$-wave multipole $E_{0+}$ (see Fig.~\ref{fig:multipole}), the solution BnGa-2014-02 did not contain any singularity corresponding to the $p \eta^{\prime}$ threshold (see also Fig.~\ref{fig:multipole}). Now, comparing both plots in Fig.~\ref{fig:beamasymLfits}, it is seen that the abrupt rise of $\left(a_{4}\right)^{\check{\Sigma}}_{4}$ is described much better using the multipoles of the more recent BnGa solution. The PWA curve shows a pronounced cusp itself at the $p \eta^{\prime}$ threshold (see black solid curve) that is sensitive to the $\left<S,G\right>$ interference term. The observed systematic deviation of the $\left(a_{4}\right)^{\check{\Sigma}}_{4}$ data to the BnGa-2019 solution is probably caused by a too small $G$-wave contribution in the $p\eta$ channel and due to the nature of the multichannel fit of the BnGa PWA.

To reconstruct the $S$ wave $E_{0+}$ as an analytic function in the complex energy plane, the correct implementation of all its analytic properties is of vital importance, i.e., the knowledge about the position and nature of the occurring singularities. The $p\eta'$ cusp observed in our data for the beam asymmetry $\Sigma$ now enforces the existence of a branch point right at the $p \eta'$ threshold. If this branch point and thus the $p \eta'$ channel 
is not implemented in a PWA, it implies that the remaining singularities of the $p\eta$ $S$ wave have to be affected. In particular, this includes the positions and residues of the resonance poles included in the PWA. As an illustration, we point to the properties of the resonance $N (1895) \frac{1}{2}^{-}(S_{11})$, which is required in the BnGa and the most recent solution of the $\eta$MAID PWA (cf. Table \ref{tab:BReta}). In both cases, the $p \eta'$ cusp is implemented and the $N (1895) \frac{1}{2}^{-}(S_{11})$ resonance is reported with very similar resonance masses \cite{etapaper,Tiator:2018heh}. 
However, the most recent solution of the J{\"{u}}Bo PWA, J\"{u}Bo-2017, which includes the CLAS beam asymmetry data but does {\it not} include the $p \eta'$  channel, the resonance $N (1895) \frac{1}{2}^{-}(S_{11})$ is not even contained. The inclusion or noninclusion of the $p \eta'$ cusp also becomes apparent when plotting the $S$ waves of BnGa, $\eta$MAID, and J\"{u}Bo against each other (see Fig.~\ref{fig:multipole}). In particular, the J\"{u}Bo solution is missing the sudden drop in the curve for $\left| E_{0+} \right|^{2}$ at the $p\eta'$ threshold. 

\begin{table}
\centering
\begin{tabular}{|c|c|c|c|c|}
\hline
Res. & & $M_{\text{pole/BW}}$ & $\Gamma_{\text{pole/BW}}$& BR($N^*\to N\eta$) \\
\hline
$N(1535)\frac{1}{2}^-$ & \scriptsize{BnGa-2019} & \scriptsize{$1496\pm4$} & \scriptsize{$125\pm6$} & \scriptsize{$0.41\pm 0.04$} \\
$(S_{11})$ & \scriptsize{J\"uBo-2017} & \scriptsize{$1495\pm 2$} & \scriptsize{$112\pm 1$} & \scriptsize{$0.64\pm 0.02$}\\
 & \scriptsize{$\eta$MAID-2018} & \scriptsize{$1522\pm 8$} & \scriptsize{$175 \pm 25$} & \scriptsize{$0.34 \pm 0.05$}\\
& \scriptsize{PDG (pole)} & \scriptsize{1500 - 1520} & \scriptsize{110 - 150} & \multirow{2}{*}{\scriptsize{0.30 - 0.55}} \\
& \scriptsize{PDG (BW)} & \scriptsize{1515 - 1545} & \scriptsize{125 - 175} &  \\
\hline
$N(1650)\frac{1}{2}^-$& \scriptsize{BnGa-2019} & \scriptsize{$1664\pm4$} & \scriptsize{$98\pm6$} & \scriptsize{$0.33\pm 0.04$} \\
$(S_{11})$& \scriptsize{J\"uBo-2017} & \scriptsize{$1674\pm 3$} & \scriptsize{$130\pm 9$} & \scriptsize{$0.07\pm 0.02$} \\
& \scriptsize{$\eta$MAID-2018} & \scriptsize{$1626^{+10}_{-5}$} & \scriptsize{$133\pm 20$} & \scriptsize{$0.19 \pm 0.06$}\\
 & \scriptsize{PDG (pole)} & \scriptsize{1640 - 1670} & \scriptsize{100 - 170} & \multirow{2}{*}{\scriptsize{0.15 - 0.35}}\\
& \scriptsize{PDG (BW)} & \scriptsize{1635 - 1665} & \scriptsize{100 - 150} & \\
\hline
$N(1895)\frac{1}{2}^-$& \scriptsize{BnGa-2019} & \scriptsize{$1907\pm 10$} & \scriptsize{$100^{+40}_{-10}$} & \scriptsize{$0.10\pm 0.05$}\\
$(S_{11})$ & \scriptsize{J\"uBo-2017} & \scriptsize{not seen} & \scriptsize{-} & \scriptsize{-} \\
& \scriptsize{$\eta$MAID-2018} & \scriptsize{$1894.4^{+5}_{-15}$} & \scriptsize{$71^{+25}_{-13}$} & \scriptsize{$0.033 \pm 0.015$} \\
 & \scriptsize{PDG (pole)} & \scriptsize{1890 - 1930} & \scriptsize{80 - 140} & \multirow{2}{*}{\scriptsize{0.15 - 0.40}}\\
& \scriptsize{PDG (BW)} & \scriptsize{1870 - 1920} & \scriptsize{80 - 200} & \\
\hline
\hline
$N(2190)\frac{7}{2}^-$& \scriptsize{BnGa-2019} & \scriptsize{$2140\pm20$} & \scriptsize{$420^{+120}_{-40}$} & \scriptsize{$0.04\pm 0.02$}\\
$(G_{17})$& \scriptsize{J\"uBo-2017} & \scriptsize{$2084\pm 7$} & \scriptsize{$281\pm 6$} & \scriptsize{$0.001\pm 0.001$} \\
& \scriptsize{$\eta$MAID-2018} & \scriptsize{2250} & \scriptsize{591} & \scriptsize{0.0453} \\
 & \scriptsize{PDG (pole)} & \scriptsize{2050 - 2150} & \scriptsize{300 - 500} & \multirow{2}{*}{\scriptsize{0.01 - 0.03}} \\
& \scriptsize{PDG (BW)} & \scriptsize{2140 - 2220} & \scriptsize{300 - 500} &  \\
\hline
$N(2250)\frac{9}{2}^-$& \scriptsize{BnGa-2019} & \scriptsize{$2195\pm 45$} & \scriptsize{$470\pm 50$} & \scriptsize{-} \\
$(G_{19})$& \scriptsize{J\"uBo-2017} & \scriptsize{$1910\pm 53$} & \scriptsize{$243\pm 73$} & \scriptsize{$0.023\pm  0.006$} \\
& \scriptsize{$\eta$MAID-2018} & \scriptsize{2250} & \scriptsize{733} & \scriptsize{0.0349}\\
 & \scriptsize{PDG (pole)} & \scriptsize{2150 - 2250} & \scriptsize{350 - 500} & \multirow{2}{*}{\scriptsize{-}}\\
& \scriptsize{PDG (BW)} & \scriptsize{2250 - 2320} & \scriptsize{300 - 600} & \\
\hline
\end{tabular}
\caption{Overview of the properties of all resonances (Res.) contributing to the $S$ and $G$ waves listed for three different PWA solutions (BnGa \cite{etapaper}, J\"uBo \cite{JuBo17}, and $\eta$MAID \cite{Tiator:2018heh}). The pole masses $M_{\text{pole}}^∗ = \text{Re}[W_{\text{pole}}]$ and widths 
$\Gamma_{\text{pole}} = -2\text{Im}[W_{\text{pole}}]$ are given in the unit MeV and the branching ratios (BR) for $N^*\to N\eta$ decays as calculated at their pole positions. Note: The values of the $\eta$MAID solution correspond to Breit-Wigner (BW) parameters. The Particle Data Group (PDG)-2019 values \cite{PDG} are provided as well. 
}
%
\label{tab:BReta}
\end{table} 

In case the $S$ wave is implemented incorrectly, this will then, via the interference terms present in Legendre moments [see, for instance, Eq.~\eqref{eq:SigmaCoeff4Composition}], also affect the other $p\eta$ multipoles since the $S$ wave is one of the most important waves in the $p\eta$ final state. It is striking that the J\"uBo-2017 PWA does not find evidence for the $N(1895)\frac{1}{2}^-(S_{11})$ resonance without the inclusion of the $p\eta'$ cusp but instead results in a high $G$-wave contribution near the $p\eta'$ threshold and finds a relatively low mass of 1910~MeV for the four star $N(2250)\frac{9}{2}^-(G_{19})$ resonance. \newline

In conclusion, we stress the importance of high-precision measurements of polarization observables with full angular coverage like the presented beam asymmetry data in the $p\eta$ final state, that allow one to observe singularities of scattering amplitudes  such as cusp effects and help to obtain a correct partial wave extraction. In particular, the existence of the $N(1895)\frac{1}{2}^- (S_{11})$ resonance has been confirmed, and precise resonance parameters have been deduced. \newline

%
We thank the technical staff of ELSA and the participating institutions for their invaluable contributions to the success of the experiment. We also thank L. Tiator, V. Kashevarov, and M. D\"oring for the fruitful discussions. We acknowledge support from the Deutsche Forschungsgemeinschaft (SFB/TR16 and SFB/TR110), the Schweizerischer Nationalfonds and the U.S. Department of Energy, Office of Science, Office of Nuclear Physics, under Contract No. DE-FG02-92ER40735.

%


\end{document}